\documentstyle[12pt]{article}
\begin{document}

{\LARGE \bf George Boole and} \\

{\LARGE \bf the Bell inequalities} \\ \\

{\bf Elem\'{e}r E ~Rosinger} \\
{\small \it Department of Mathematics \\ and Applied Mathematics} \\
{\small \it University of Pretoria} \\
{\small \it Pretoria} \\
{\small \it 0002 South Africa} \\
{\small \it eerosinger@hotmail.com} \\ \\

{\bf Abstract} \\

As shown by Pitowsky, the Bell inequalities are related to certain classes of probabilistic
inequalities dealt with by George Boole, back in the 1850s. Here a short presentation of this
relationship is given. Consequently, the Bell inequalities can be obtained {\it without} any
assumptions of physical nature, and merely through mathematical argument. \\ \\

{\bf 0. Preliminary remarks} \\

The importance of the typically quantum phenomenon of {\it entanglement} is well known. This
issue of entanglement has been, and still is of special focus in Quantum Mechanics, not least
due to its intimate connection to such fundamental disputes as {\it locality} versus {\it
nonlocality}, and the related literature is indeed vast. It is also of major importance in
the fast emerging theory of Quantum Computation. \\
The first time an entanglement based argument was used in a critically important manner in
Quantum Theory was in the celebrated EPR paper. \\

Nearly three decades after the EPR paper had appeared in 1935, John Bell published in 1964
what amounted to a surprising {\it conflict} between predictions of a classical world view
based on the principle of locality, and on the other hand, those of Quantum Mechanics. The
classical world view based on locality led J Bell to certain inequalities which, however,
proved to be {\it contradicted} by Quantum Mechanics, namely, by certain properties of
suitably chosen entangled EPR pairs. \\
And this contradiction could be observed in effective quantum mechanical experiments, such as
conducted for instance in 1982 by A Aspect et.al., see Maudlin. \\

Here it should be mentioned again that, as often in science, the related terminology which
entered the common use tends to misplace the focus. Indeed, the main point in J Bell's
contribution is not about inequalities, but about the fact that they lead to the mentioned
contradiction. Furthermore, there are by now a number of other similar arguments which all
lead to such contradictions with Quantum Mechanics. \\

Needless to say, it is well known that ever since its very inception in the 1920s, Quantum
Mechanics has been witnessing an ongoing foundational controversy related to its
interpretation, some of the earlier major stages of this controversy being those between N
Bohr and A Einstein. However, as not seldom in such human situations, a certain saturation,
stationarity and loss of interest may set in after some longer period of time has failed to
clarify enough the issues involved. \\

The surprising result of J Bell happened to appear after most of the founding fathers of
Quantum Mechanics had left the scene, and proved to inaugurate a fresh line of controversies,
see Bell, Cushing \& McMullin, Maudlin. \\

Here, an attempt is presented to recall in short the essential aspects of J Bell's result.
Clearly, at least to the extent that this result is essentially connected to the typically
quantum phenomenon of entanglement, it may be expected to be relevant for a better
understanding, and thus further development of booth Quantum Theory and Quantum Computation. \\

Also, a relatively less well know aspect of Bell type inequalities is presented here, namely
that, these inequalities are among a larger class of {\it purely probabilistic} inequalities,
a class whose study was started by George Boole, with the first results published in the
appendix of his book "The Laws of Thought", back in 1854. This purely mathematical study of
the respective inequalities was later further extended in the work of a number of
mathematicians and probabilists, see details in Pitowsky. \\

Needless to say, this fact does in no way detract from the importance and merit of J Bell's
result. Indeed, unlike J Bell, it is obvious that G Boole and his mentioned followers,
including certain probabilists in more recent times, did not consider the quantum mechanical
implications of such inequalities. In this way, the importance and merit of J Bell's result is
to single out for the first time certain rather simple inequalities which are supposed to be
universally valid, provided that a classical setup and locality are assumed, and then show
that the respective inequalities do to a quite significant extent conflict with Quantum
Mechanics, involving in this process such important issues as entanglement and locality versus
nonlocality. \\

There is a special interest in pointing out the fact that the Bell type inequalities can be
established by a {\it purely mathematical argument}, as was done, for instance, by the
followers of G Boole. In this regard it is important to note that both in the work of J Bell,
as well as in the subsequent one of many of the physicists who dealt with this issue, the true
nature of such inequalities is often quite obscured by a complicated mix of physical and
mathematical argument. Such an approach, however, is unnecessary, and can of course create
confusions about the genuine meaning, scope and implications of J Bell's result. \\

The fact however is that regardless of the considerable generality of the framework underlying
such inequalities, and thus of the corresponding minimal conditions required on locality, one
can nevertheless obtain the respective inequalities through purely mathematical argument, and
{\it without} any physical considerations involved, yet they turn out even to be testable {\it
empirically}. And in a surprising manner, they fail tests which are of a quantum mechanical
nature. And this failure is both on theoretical and empirical level. In other words, the Bell
inequalities contradict theoretical consequences of Quantum Mechanics, and on top of that,
they are also proven wrong in quantum mechanical experiments such as those conducted by
Aspect et.al. \\

The impact of Bell's inequalities is only increased by the fact that they require such {\it
minimal} conditions, yet they deliver a clear cut and unavoidable {\it conflict} with Quantum
Mechanics. \\

Let us also note the following. J Bell, when obtained his inequalities, he was concerned with
the issue of the possibility, or otherwise, of the so called {\it deterministic, hidden
variable} theories for Quantum Mechanics. This issue arose from the basic controversy in the
interpretation of Quantum Mechanics, and aimed to eliminate the probabilistic aspects involved
typically in the outcome of measurements. One way in this regard was to consider Quantum
Mechanics {\it incomplete}, and then add to it the so called hidden variables, thus making the
theory deterministic by being able a priori to specify precisely the measurement results. \\

By the way, the very title of the EPR paper was raising the question whether Quantum Mechanics
was indeed complete, and suggested the experiment with entangled quantum states in order to
justify that questioning. \\

Regarding the term hidden variables, once again we are faced with a less than proper
terminology. Indeed, as it is clear from the context in which this term has always been used,
one is rather talking about {\it missing variables}, or perhaps variables which have been
missed, overlooked or disregarded, when the theory of Quantum Mechanics was set up. Details in
this regard can be found in Holland, where an account of the de Broglie-Bohm causal approach
to Quantum Mechanics is presented. \\

In view of this historical background, the effect of Bell's inequalities is often {\it wrongly}
interpreted as proving that a deterministic hidden variable theory which is subjected to the
principle of locality is not possible. \\

However, it is important to note that such a view of Bell's inequalities is {\it not} correct.
Indeed, by giving up determinism, or the hidden variables, one still remains with Bell's
inequalities, since these inequalities {\it only} assume a classical framework in which the
locality principle holds. \\ \\

{\bf 1. Boole Type Inequalities} \\

In his mentioned book G Boole was concerned among others with conditions on all possible
experience or experimentation, this being the factual background to logic and the laws of
thought. Needless to say, G Boole assumed automatically a classical and non-quantum context
which was further subjected to the principle of locality. \\

Here we shall limit ourselves to a short presentation of some of the relevant aspects. Let
therefore $A_1, ~.~.~.~ , A_n$ be arbitrary $n \geq 2$ events, and for
$1 \leq i_1 < i_2 < ~.~.~.~ < i_k \leq n$, let $p_{i_1, i_2,  ~.~.~.~ , i_k}$ be the
probability of the simultaneous event
$A_{i_1} \bigcap A_{i_2} \bigcap ~.~.~.~ \bigcap A_{i_k}$. \\

One of the questions G Boole asked was as follows. Suppose that the only information we have
are the probabilities $p_1,~ p_2, ~.~.~.~ , p_n$ of the respective individual events
$A_1, ~.~.~.~ , A_n$. What are under these conditions on information the best possible
estimates for the probabilities of
$A_1 \bigcup A_2 \bigcup ~.~.~.~ \bigcup A_n$ and $A_1 \bigcap A_2 \bigcap ~.~.~.~
\bigcap A_n$ ? \\

G Boole gave the following answers which indeed are correct

\bigskip
(1.1) \quad $ \begin{array}{l} \mbox{max}~ \{~ p_1,~ p_2, ~.~.~.~ , p_n ~\} ~\leq~ P (  A_1 \bigcup A_2 \bigcup ~.~.~.~
                                                    \bigcup A_n ) ~\leq~ \\ \\
                                                        ~~~~~~~~~~~~~~~~~~~~~~~~~~~~~~~~~~~~~~~~~~\mbox{min}~ \{~ 1,~ p_1 + p_2 + ~.~.~.~  + p_n ~\}
                        \end{array} $ \\

\medskip
(1.2) \quad $ \begin{array}{l} \mbox{max}~ \{~ 0,~ p_1 + p_2 + ~.~.~.~ + p_n -n + 1 ~\} ~\leq~ \\ \\
                                                  ~~~~~~~~~\leq~ P (  A_1 \bigcap A_2 \bigcap  ~.~.~.~ \bigcap A_n ) ~\leq~
                                                                                            ~\mbox{min}~ \{~  p_1 ,~ p_2, ~.~.~.~  , p_n ~\}
                         \end{array} $ \\

\medskip
And these are the best possible inequalities in general, since for suitable particular cases
equality can hold in each of the four places. \\

A rather general related result is the so called {\it inclusion-exclusion principle} of Henri
Poincar\'{e}

\bigskip
(1.3) \quad $ \begin{array}{l} P (  A_1 \bigcup A_2 \bigcup ~.~.~.~  \bigcup A_n ) ~=~ \Sigma_{1 \leq i \leq n}~ p_i ~-~
                                                  \Sigma_{1 \leq i < j \leq n}~ p_{i j} ~+~ \\ \\
                                      ~~~~~~~~~~~~~~~+~ \Sigma_{1 \leq i < j < k \leq n}~ p_{i j k} ~+~ ~.~.~.~  +~ ( - 1 )^{n + 1}~ p_{1 2 ~.~.~. n}

                        \end{array} $ \\

\medskip
This however requires the knowledge of the probabilities of all the simultaneous events
$A_{i_1} \bigcap A_{i_2} \bigcap  ~.~.~.~ \bigcap A_{i_k}$, with
$1 \leq i_1 < i_2 < ~.~.~.~ < i_k \leq n$. \\

A question with less demanding data, yet with more of them than required in (1.1) and
(1.2), is the following. Suppose we know the probabilities $p_i$ of the events $A_i$, with
$1 \leq i \leq n$, as well as the probabilities $p_{i j}$ of the simultaneous events
$A_i \bigcap A_j$, with $1 \leq i < j \leq n$. \\
What is then the best possible estimate for the probability of \\
$A_1 \bigcup A_2 \bigcup ~.~.~.~  \bigcup A_n$ ? \\

Unfortunately, this question is computationally intractable, Pitowsky. However, C E Bonferroni
gave some answers in 1936, one of which is that

\bigskip
(1.4) \quad $ \Sigma_{1 \leq i \leq n}~ p_i ~-~ \Sigma_{1 \leq i < j \leq n}~ p_{i j} ~\leq~ P (  A_1 \bigcup A_2 \bigcup ~.~.~.~
                                                    \bigcup A_n ) $ \\

\medskip
and here it is interesting to note that (1.4) generates easily $2^n - 1$ other independent
inequalities by the following procedure. We take any
$1 \leq i_1 < i_2 < ~.~.~.~ < i_k \leq n$, and replace in (1.4) the events $A_{i_l}$, for
$1 \leq l \leq k$, with their complementary sets. \\

Now the important fact to note is that Bell's inequalities result from (1.4) in this way, in
the case of $n = 3$. \\

It follows therefore that Bell's inequalities are of a purely mathematical nature, and as such,
only depend on classical probability theory. \\

By the way, Boole's inequalities and its further developments have been presented in well
known monographs of mathematics and probability theory, some of them as recently as in 1970,
and related research has continued in mathematics and in probability theory till the present
day, Pitowsky. As so often however, due to extreme specialization and the corresponding
narrowing of interest, such results seem not to be familiar among quantum physicists. In this
regard it may be worth mentioning that Pitowsky himself is a philosopher of science. \\ \\

{\bf 2. The Bell Effect} \\

There are by now known a variety of ways which describe the phenomenon brought to light for
the first time by Bell's inequalities. In order to avoid complicating the issues involved, we
shall present here one of the most simple such ways, Maudlin. \\
This phenomenon, which one can call the {\it Bell effect} is a {\it contradiction} resulting
between Quantum Mechanics, and on the other hand, what can be done in a classical setup which
satisfies the principle of locality. The Bell inequalities are only one of the ways, and
historically the first, which led to such a contradiction. They will be presented in section
4. What is given here is a simple and direct argument leading to the mentioned kind of
contradiction. \\

Certain {\it entangled} quantum particles can exhibit the following behaviour. After they
become spatially separated, they each can be subjected to three different experiments, say, A,
B and C, and each of them can produce one and only one of two results, which for convenience
we shall denote by R and S, respectively. \\
What is so uniquely specific to these entangled quantum particles is the behaviour described
in the next three conditions which such particles do satisfy. \\

{\it Condition 1}. When both particles are subjected to the same experiment, they give the
same result. \\

{\it Condition 2}. When one of the particles is subjected to A and the other to B, or one is
subjected to B and the other to C, they will in a large number of experiments give the same
result with a frequency of 3/4. \\

{\it Condition 3}. When one of the particles is subjected to A and the other to C, then in a
large number of experiments they will give the same result with a frequency 1/4. \\

Now, the surprising fact is that {\it no} experiment in a classical setup in which the
principle of locality holds can come anywhere near to such a behaviour. \\
And strangely enough, that includes as well the case when two conscious participants, and not
merely two physical entities would be involved. In such a case, when conscious participant are
present, we shall see the experiments A, B and C as questions put to the two participants,
while the results R and C will be seen as their respective answers. \\

Such are indeed the wonders of {\it entanglement} and of certain EPR pairs that some of their
performances, like for instance those which satisfy conditions 1, 2 and 3 above, {\it cannot}
be reproduced in a classical context which obeys the locality principle, even if attempted by
two conscious participants. \\

Indeed, a simple analysis shows that the best two such conscious participants can do is to
decide to give the same answers, when asked the same questions. This means that any possible
{\it strategy} of the two participants has to be {\it joint} or identical, and as such, it is
given by a function \\

$~~~ f : \{~ A, B, C ~\} \longmapsto \{~ R, S ~\} $ \\

Clearly, there are 8 such joint strategies, namely

\bigskip
(2.1) \quad $ \begin{array}{l} ~~~~~~~~~ A ~~~~~~ B ~~~~~~ C \\
                                                  ---------- \\ \\
                                                  1 ~~~~~~~ R ~~~~~~ R ~~~~~~ R \\
                                                  2 ~~~~~~~ R ~~~~~~ R ~~~~~~ S \\
                                                  3 ~~~~~~~ R ~~~~~~ S ~~~~~~ R \\
                                                  4 ~~~~~~~ R ~~~~~~ S ~~~~~~ S \\
                                                  5 ~~~~~~~ S ~~~~~~ R ~~~~~~ R \\
                                                  6 ~~~~~~~ S ~~~~~~ R ~~~~~~ S \\
                                                  7 ~~~~~~~ S ~~~~~~ S ~~~~~~ R \\
                                                  8 ~~~~~~~ S ~~~~~~ S ~~~~~~ S
                        \end{array} $ \\

\medskip
Now it is obvious that by choosing only these 8 join strategies, condition 1 above will be
satisfied. \\
From the point of view of satisfying conditions 2 and 3 above, the strategy pairs ( 1, 8 ),
( 4, 5 ), ( 3, 6 ) and ( 2, 7 ) are equivalent. Therefore, we only remain with four distinct
strategies to consider, namely

\bigskip
(2.2) \quad $ \begin{array}{l} ~~~~~~~~~ A ~~~~~~ B ~~~~~~ C \\
                                                  ---------- \\ \\
                                                  1 ~~~~~~~ R ~~~~~~ R ~~~~~~ R \\
                                                  2 ~~~~~~~ R ~~~~~~ R ~~~~~~ S \\
                                                  3 ~~~~~~~ R ~~~~~~ S ~~~~~~ R \\
                                                  4 ~~~~~~~ R ~~~~~~ S ~~~~~~ S \\
                         \end{array} $ \\

\medskip
At this point the two participants can further improve on their attempt to satisfy conditions
2 and 3 above by randomizing their joint strategies. For that purpose, they can choose four
real numbers $\alpha, \beta, \gamma, \delta \in {\bf R}$, such that

\bigskip
(2.3) \quad $ \begin{array}{l} \alpha,~ \beta,~ \gamma,~ \delta ~\geq~ 0 \\ \\
                                                  \alpha ~+~\beta ~+~ \gamma ~+~ \delta ~=~ 1
                        \end{array} $ \\

\medskip
and use their joint strategies 1, 2, 3 and 4 with the respective frequencies $\alpha, \beta,
\gamma, \delta$. A simple computation will show that conditions 2 and 3 above will further
impose on $\alpha, \beta, \gamma, \delta$ the relations

\bigskip
(2.4) \quad $ \begin{array}{l} \gamma ~+~ \delta ~=~ 1/4 \\ \\
                                                  \beta ~+~ \gamma ~=~ 1/4 \\ \\
                                                  \beta ~+~ \delta ~=~ 3/4
                        \end{array} $ \\

\medskip
However, (2.3) and (2.4) yield

\bigskip
(2.5) \quad $ \gamma ~=~ - 1/8 $ \\

\medskip
thus a {\it contradiction}. \\

Here it is important to note that the {\it locality} principle was assumed in (2.1) - (2.5).
In other words, each of the two participants could be asked questions, without the question
asked from one of them having any effect on the answer of the other. Indeed, the two
participants could be asked different questions, and each of them would only reply according
to the question asked, and according to their joint strategy, which they happened to use at
the moment. \\

The fact that the setup in (2.1) - (2.5) is {\it classical}, that is, it is not specifically
quantum mechanical, is obvious. \\ \\

{\bf 3. Bell's Inequalities} \\

For convenience we shall consider two entangled quantum particles which are in a situation
even simpler than in section 3, Cushing \& McMullin. Namely, each of the particles can only be
subjected to two different experiments, and as before, each such experiment can only give one
of two results. \\
In view of the specific quantum mechanical setup considered, the experiments to which the two
particles are subjected can be identified with certain angles in $[ 0, 2 \pi ]$ which define
the directions along which quantum spins are measured. As far as the results obtained, they
can be identified with quantum spins, and as such will be denoted by $+$ and $-$, respectively.
Finally, when the same experiment is performed on both particles, it is assumed that due to
their entanglement and momentum conservation, the results are always different, that is, one
result is $+$, while the other is $-$. \\

Locality, as before, will mean that, when far removed in space from one another, each particle
can be subjected to any experiment independently, and the result does not depend on what
happens with the other particle. \\

Having done a large number of experiments on such two particles, let us denote by

\bigskip
(3.1) \quad $ p_{1, 2}~ ( \alpha_i, \beta_j ~|~ x, y ) $ \\

\medskip
the probability that experiment $\alpha_i \in [ 0, 2 \pi ]$, with $i \in \{~ 1, 2 ~\}$, on
particle 1 yields the result $x \in \{~ +, - ~\}$, and at the same time experiment $\beta_j
\in [ 0, 2 \pi ]$, with $j \in \{~ 1, 2 ~\}$, on particle 2 yields the result $y \in
\{~ +, - ~\}$. \\
Similarly we denote by

\bigskip
(3.2) \quad $ p_1~ ( \alpha_i ~|~ x ),~~~ p_2~ ( \beta_j ~|~ y ) $ \\

\medskip
the respective probabilities that experiment $\alpha_i \in [ 0, 2 \pi ]$, with $i \in
\{~ 1, 2 ~\}$, on particle 1 yields the result $x \in \{~ +, - ~\}$, and that experiment
$\beta_j \in [ 0, 2 \pi ]$, with $j \in \{~ 1, 2 ~\}$, on particle 2 yields the result
$y \in \{~ +, - ~\}$. \\

Now based alone on the assumption of locality, one obtains {\it Bell's inequality}

\bigskip
(3.3) \quad $ \begin{array}{l} - 1 ~\leq~ p_{1, 2}~ ( \alpha_1, \beta_1 ~|~ +, + ) ~+~ p_{1, 2}~ ( \alpha_1, \beta_2  ~|~ +, + ) ~+~ \\ \\
                       ~~~~~~ ~+~ p_{1, 2}~ ( \alpha_2, \beta_2 ~|~ +, +  ) ~-~ p_{1, 2}~ ( \alpha_2, \beta_1 ~|~ +, + ) ~-~ \\ \\
                       ~~~~~~ ~-~ p_1~ ( \alpha_1 ~|~ + ) ~-~ p_2~ ( \beta_2 ~|~ + ) ~\leq~ 0
                 \end{array} $ \\

\medskip
Obviously, by changing the indices of the angles and the spin values, one can obtain further
variations of this inequality. \\

What suitable quantum mechanical experiments can give are very good approximations of the
relations

\bigskip
(3.4) \quad $ \begin{array}{l} p_{1, 2}~ ( \alpha, \beta ~|~ +, + ) ~=~ p_{1, 2}~ ( \alpha, \beta ~|~ -, - ) ~=~
                                         ( 1 / 2 ) \sin^2 ( \alpha - \beta ) / 2 \\ \\
                                 p_{1, 2}~ ( \alpha, \beta ~|~ +, - ) ~=~ p_{1, 2}~ ( A, B ~|~ -, + ) ~=~
                                         ( 1 / 2 ) \cos^2 ( \alpha - \beta ) / 2 \\ \\
                                 p_1~ ( \alpha ~|~ + ) ~=~ p_2~ ( \beta ~|~ - ) ~=~ 1 / 2
                 \end{array} $ \\

\medskip
where $\alpha, \beta \in [ 0, \pi ]$. \\

Now let us return to the Bell inequality in (3.3) and take following angles for the
experiments

\bigskip
(3.5) \quad $ \alpha_1 ~=~ \pi / 3,~~~ \alpha_2 ~=~ \pi,~~~ \beta_1 ~=~ 0,~~~ \beta_2 ~=~ 2 \pi /3 $ \\

\medskip
in which case we obtain the {\it contradiction}

\bigskip
(3.6) \quad $ - 1 / 8 ~\geq~ 0 $ \\

\medskip
As shown in Pitowsky, the Bell inequality in (3.3), as well as its mentioned variants follow
from the Bonferroni inequalities in (3). \\

Let us conclude the issue of Bell's inequalities, and more importantly, of the Bell Effect, by
noting that the resulting contradictions show the existence of relevant physics {\it beyond}
any classical framework which obeys the principle of locality. \\
And the quantum mechanical experiments which, together with Bell's inequalities, deliver the
above contradictions are therefore part of such a physics, even if Quantum Mechanics as a
theory is still quite far from having at last settled its foundational controversies. \\

As far as {\it entangled} quantum particles, or in general, systems are concerned, they are
some of the simplest quantum phenomena to lead to the Bell Effect, and thus beyond the
classical and local framework. This is therefore one of the reasons why they can offer
possibilities in quantum computation which cannot be reached anywhere near by usual electronic
digital computers, which obviously belong to realms of physics that are classical and
subjected to the locality principle. \\ \\

{\bf 4. Locality versus Nonlocality} \\

The original EPR paper, then the de Broglie-Bohm causal interpretation, as well as Bell's
inequalities have focused a considerable attention on the issue of locality versus nonlocality.
And in view of what appear to be obvious reasons, there is a rather unanimous and strong
dislike of nonlocality among physicists. A typical instance of such a position is illustrated
by the next citation from a letter of A Einstein to Max Born, see Maudlin, or Born :

\bigskip
\begin{quote}
... If one asks what, irrespective of quantum mechanics, is characteristic of the world of
ideas of physics, one is first of all struck by the following : the concepts of physics relate
to a real outside world, that is, ideas are established relating to things such as bodies,
fields, etc., which claim "real existence" that is independent of the perceiving subject -
ideas which, on the other hand, have bee brought into as secure a relationship as possible
with the sense-data. It is further characteristic of these physical objects that they are
thought of as arranged in a space-time continuum. An essential aspect of this arrangement of
things in physics is that they lay claim, at a certain time, to an existence independent of
one another, provided these objects "are situated in different parts of space". Unless one
makes this kind of assumptions about the independence of the existence (the "being-thus") of
objects which are far apart from one another in space - which stems in the first place from
everyday thinking - physical thinking in the familiar sense would not be possible. It is also
hard to see any way of formulating and testing the laws of physics unless one makes a clear
distinction of this kind. This principle has been carried to extremes in the field theory by
localizing the elementary objects on which it is based and which exist independently of each
other, as well as the elementary laws which have been postulated for it, in the infinitely
small (four dimensional) elements of space. \\
The following idea characterizes the relative independence of objects far apart in space (A
and B) : external influence on A has no direct influence on B; this is known as the "principle
of contiguity", which is used consistently in the field theory. If this axiom were to be
completely abolished, the idea of laws which can be checked empirically in the accepted sense,
would become impossible...
\end{quote}

\medskip
However, as often happens in the case of strongly felt dislikes, the reactions involved may
prove to be exaggerated. And in the case of nonlocality this seems to happen. \\
Indeed, certain milder, fast diminishing forms of nonlocality have been around in physics, and
some of them, like the gravitational effect of a mass, were introduced by no lesser
contributors than Isaac Newton. Of course, the gravitational effect of a given mass, although
it decreases fast, namely, with the square of the distance, it is nevertheless not supposed to
vanish completely anywhere. A similar thing is supposed to happen with the electric charge,
according to Culomb's law. \\
On the other hand, certain nonlocality effects in the case of entangled quantum particles are
{\it not} supposed to diminish at all with the distance separating the particles. \\

What seems to happen, however, is that there is a significant reluctance to admit even one
single, and no matter how narrow and well circumscribed instance of a {\it nondiminishing
nonlocality}. Such a reluctance appears to be based on the perception that the acceptance of
even one single such nondiminishing nonlocality would instantly bring with it the collapse of
nearly all of the theoretical body of physics. \\

In other words, it is considered that physical theory, as it stands, is {\it critically
unstable} with respect to the incorporation of even one single nondiminishing nonlocality. \\

Clearly, if indeed such may be the case, then that should rather be thoroughly investigated,
instead of being merely left to perceptions as part of an attitude which, even if by default,
treats it as a taboo. \\
After all, a somewhat similar phenomenon was still going on less than four centuries ago, when
the idea of Galileo that our planet Earth is moving was felt to be an instant and mortal
threat to the whole edifice of established theology. \\ \\

\end{document}